# Reducing post-surgery recovery bed occupancy with a probabilistic forecast model

Belinda Spratt[1] and Erhan Kozan

School of Mathematical Sciences, Queensland University of Technology, Brisbane, AU

Operations Research approaches to surgical scheduling are becoming increasingly popular in both theory and practice. Often these models neglect stochasticity in order to reduce the computational complexity of the problem. We wish to provide practitioners and hospital administrative staff with a start-of-day probabilistic forecast for the occupancy of post-surgery recovery spaces. The model minimises the maximum expected occupancy of the recovery unit, thus levelling the workload of hospital staff, and reducing the likelihood of bed shortages and bottlenecks. We show that a Poisson binomial random variable models the number of patients in the recovery when parameterised by the surgical case sequence. A mixed integer nonlinear programming model for the surgical case sequencing problem reduces the maximum expected occupancy in post-surgery recovery spaces. Simulated Annealing produces good solutions in short amounts of computational time. We evaluate the methodology with a full year of historical data. The solution techniques reduce maximum expected recovery occupancy by 18% on average. This alleviates a large amount of stress on staff in the post-surgery recovery spaces, reduces the likelihood of bottlenecks, and improves the quality of care provided to patients.

[1] corresponding author:

b.spratt@qut.edu.au +61731383035

# 1 Introduction

In recent years, academics and practitioners have placed a great deal of focus on improving the performance of the operating theatre (**OT**) through use of Operations Research. Given an increasing number of surgical requests, and a fixed number of resources, it is necessary to increase surgical throughput in order to keep up with demand. Practitioners often schedule the OT as a stand-alone entity, despite potential impacts on the rest of the hospital.

Surgical schedules can have a large impact on downstream wards (e.g. patient recovery spaces). Not only is it necessary to ensure that patient demand on downstream wards does not exceed capacity, but it is also necessary to ensure that there are sufficient staff available at times of peak demand. As such, we present a probabilistic forecast for recovery bed occupancy based on the assumption of lognormally distributed surgery and recovery durations. The forecast uses only start-of-day information.

Using this probabilistic forecast, we propose a Mixed Integer Nonlinear Programming (**MINLP**) formulation of the Surgical Case Sequencing Problem (**SCSP**) where the objective is to minimise the maximum expected occupancy level. We solve this model using a novel constructive heuristic and Simulated Annealing (**SA**). This approach improves the surgical case sequences (**SCSs**) produced by a large Australian public hospital in terms of lowering downstream occupancy levels without reducing throughput or utilisation of the OT.

The contribution of this paper is two-fold. Firstly, we present an analytical model for the distribution of post-surgery recovery bed occupancy based on lognormally distributed surgical and recovery durations (cf. Section 3.2). Secondly, we present an innovative MINLP formulation of the SCSP where the objective is to minimise the maximum expected recovery bed occupancy level (cf. Section 3.3). We are able to create major improvements in downstream occupancy by making minor changes to the surgical schedule by modifying the order of patients or shifting surgery start times. We show that the analytical model presented in Section 3.2 can be implemented to produce surgical sequences that are less erratic, thus reducing the stress on post-surgery recovery staff and providing more information to hospital planning and rostering staff members. The method is computationally efficient and is suitable for implementation in both planning and reactive rescheduling methodologies.

The remainder of the paper is organised as follows. In Section 3.1, we discuss the motivating case study and provide a description of the problem. Section 3.2 contains the derivation of a Poisson Binomial model of recovery bed occupancy. This leads to the MINLP formulation in Section 3.3. Due to the complexity of the problem, we implement SA to produce good feasible solutions (cf. Section 3.4). We present and discuss results in Section 4. Section 5 includes practice implications and concluding remarks.

## 2   Literature Review

OT planning and scheduling occurs at three main levels: strategic, tactical, and operational. These problems are addressed on long, medium, and short term planning horizons respectively (Cardoen, Demeulemeester, & Beliën, 2010). There is often a hierarchical approach to OT planning and scheduling. Strategic decisions influence tactical decisions which then, in turn, influence operational decisions. Furthermore, the surgical department has the highest impact on hospital workload (Vanberkel et al., 2011). For a thorough review of OT planning and scheduling, see Cardoen et al. (2010). A review of the use of Operations Research in other hospital wards is provided by van de Vrugt, Schneider, Zonderland, Stanford, and Boucherie (2018).

Whilst understanding the patient occupancy distributions caused by tactical decisions is useful in medium term staff planning, in this paper we focus on how downstream occupancy changes throughout the day. To capture the daily variability in downstream ward occupancy caused by surgical throughput, we must consider operational level OT planning and scheduling decisions. As such, we focus on the operational level of OT planning and scheduling to forecast recovery bed occupancy.

Min and Yih (2010) provide a closed form model for the expected number of patients in the surgical intensive care unit (**SICU**) based on the assumption of normally distributed surgical durations and uniformly distributed patient length of stay (**LOS**) in the SICU. This model minimises patient related costs whilst assigning patients to OT time blocks. The closed form model provided by Min and Yih (2010) is inappropriate whenever these distribution assumptions are violated. In reality, surgical durations are closer to a lognormal distribution (Strum, May, & Vargas, 1998). Recovery durations are also well-modelled by the lognormal distribution (Spratt, Kozan, & Sinnott, 2018). As such, we propose a new model to forecast post-surgical recovery bed occupancy that utilises lognormal surgical durations and recovery times.

Several authors have considered using simulation to ensure deterministic models are suitable in a stochastic environment. Chow, Puterman, Neda, Wenhai, and Derek (2011) implement a surgical scheduling model based on mixed integer programming (**MIP**) and Monte Carlo simulation to reduce peak bed occupancies. Price et al. (2011) use integer programming and discrete event simulation to create surgical block schedules and evaluate their impact on the PACU. One of the limitations of the optimisation-simulation approach used by Chow et al. (2011) and Price et al. (2011) is the computational time required to solve an IP or MIP model and perform sufficiently many simulations to understand the impacts in a stochastic environment. Lee and Yih (2014) also identify computational time as a barrier to adoption of their work scheduling surgeries with consideration of uncertain durations and PACU resources. Our introduction of an innovative constructive heuristic and use of a metaheuristic alongside an explicit form for recovery bed occupancy reduces computational time to only six seconds.

Latorre-Núñez et al. (2016) simultaneously schedule the OT and post anaesthesia recovery spaces when solving the surgical case sequencing problem (**SCSP**). When producing their schedules, the authors consider only mean durations of both surgery and recovery. This method disregards any additional information that planners and schedulers may have regarding task duration distributions. Latorre-Núñez et al. (2016) also identify the use of deterministic surgery and recovery durations as a limitation; we address this by using stochastic surgical durations in our work.

Decomposition approaches and/or assumptions around the availability of predefined schedules are common methods for patient scheduling problems. Recently, Bai, Storer, and Tonkay (2017) considered the multiple operating room (**OR**)[2] surgical scheduling problem with random surgical durations and PACU capacity constraints. The authors assume that there is a predefined patient sequence for each OR. Fairley, Scheinker, and Brandeau (2018) use gradient tree boosting to model surgery and recovery durations in a children's hospital. The authors then use two deterministic integer programming models to firstly minimise the finishing time of each OR and then to schedule procedures to level occupancy. In this work, we sequence patients and schedule patients in an integrated approach, overcoming any limitations associated with suboptimal predefined patient sequences or decomposition approaches.

---

[2] Note: The OT is the set of ORs.

# 3 Method

## 3.1 *Case Study and Problem Description*

We base the work in this paper on a case study of an Australian public hospital. The hospital has a large surgical department consisting of 21 ORs. There are around twenty bed spaces available in the surgical care unit (**SCU**) and post-anaesthesia care unit (**PACU**).

The hospital utilises a modified block scheduling policy when generating a four week, rotating master surgical schedule (**MSS**). This means that they may reallocate OR time in the case that a surgical team is unavailable or does not require a particular OR block. In this case, other specialties are able to request this time based on the patient demand. Planning staff rarely modify the MSS, and the schedule has remained largely unchanged since the hospital opened.

The hospital schedules nurses and anaesthetists more regularly based on the current MSS. When scheduling nurses and anaesthetists, the hospital must consider their preference or experience regarding the different surgical specialties. Adjustments to the MSS do not consider the availability of nurses and anaesthetists as there are usually sufficient staff available.

At the case study hospital, surgeons work in conjunction with case managers and other administrative staff members to plan and schedule the OT department. Each week, surgeons produce lists of patients that they would like to see during their allocated surgical time. The hospital holds an elective bookings meeting every Thursday to discuss potential cases in the upcoming week. In some cases, surgeons do not provide a list for a particular OR block as they expect non-elective arrivals will occur in the meantime.

On the day of surgery, elective patients are required to arrive on time for their surgery. The hospital allocates wards to inpatients (expected to stay for more than a single day). The inpatient is then sent either to their ward or to prepare in the SCU depending on the time until surgery and whether any other procedures are required beforehand.

Most patients recover from surgery in the PACU, with recovery times depending on the type of surgery. Any surgical transfers from the intensive care unit (**ICU**) return directly to the ICU following surgery. Once a patient has recovered sufficiently in the PACU, day patients continue recovery in the SCU whereas inpatients return to the ward or to the ICU. The PACU forms a bottleneck in the system due to its limited capacity. A surgery cannot proceed if there is overcrowding in the PACU. Similarly, a patient may be held in the OR until a space becomes available in the PACU. This, in turn, results in a delay in subsequent surgeries.

In the past, OT planning and scheduling at the hospital did not consider downstream resources. Here, we propose that staff sequence patients so as to reduce the strain on the PACU and create a more efficient surgical department by relieving the bottleneck in the system.

### 3.2 *Forecasting Recovery Bed Occupancy*

The first step in sequencing patients to reduce PACU strain is to understand how the sequence of patients affects the occupancy of the PACU. Spratt et al. (2018) show that both surgery and recovery durations at the case study hospital are well-modelled by the lognormal distribution.

Whilst there appear to be some deviations from the fitted lognormal distribution, we obtain a better fit when patients are further categorised by their specialty and ASA (American Society of Anaesthesiologists) physical status classification code. The ASA code identifies the health of a patient prior to their surgery and ranges from one (healthy) to six (brain-dead).

In order to forecast the occupancy of recovery beds in the PACU at the start of the day, we assume that patient surgery and recovery durations follow a lognormal distribution, with parameters dependent on surgical specialty and ASA code. The remainder of this section is organised as follows. First, we derive the probability an individual patient is in recovery at any given time. We then discuss the distribution, expectation, and variance of the total number of patients in recovery at any given time.

### 3.2.1 Individual Recovery

Here, patient $p$'s surgical duration and recovery duration are assumed to be lognormally distributed, and are denoted $S_p$ and $R_p$ respectively. We use a lognormal distribution to approximate the sum of lognormal distributions. $T_p$ is the random variable representing the total duration of patient p's surgery and recovery. To simplify these calculations, we assume that a patient's recovery time is independent of their surgery time.

$$S_p \sim \text{Lognormal}(\mu_p, \sigma_p^2)$$

$$F_{S_p}(t) = \frac{1}{2} + \frac{1}{2}\text{erf}\left(\frac{\ln(t) - \mu_p}{\sqrt{2}\sigma_p}\right), \qquad t \geq 0$$

$$R_p \sim \text{Lognormal}(\tilde{\mu}_p, \tilde{\sigma}_p^2)$$

$$T_p \approx S_p + R_p$$

$$T_p \sim \text{Lognormal}(\hat{\mu}_p, \hat{\sigma}_p^2)$$

$$F_{T_p}(t) = \frac{1}{2} + \frac{1}{2}\text{erf}\left(\frac{\ln(t) - \hat{\mu}_p}{\sqrt{2}\hat{\sigma}_p}\right), \qquad t \geq 0$$

We determine the parameter values for the random variable $T_p$ through moment matching as per Stuart and Kozan (2012).

$$\hat{\mu}_p = \ln\left(\frac{M_p^2}{\sqrt{V_p + M_p^2}}\right)$$

$$\hat{\sigma}_p^2 = \ln\left(\frac{V_p + M_p^2}{M_p^2}\right)$$

$$M_p = e^{\mu_p + \frac{\sigma_p}{2}} + e^{\tilde{\mu}_p + \frac{\tilde{\sigma}_p}{2}}$$

$$V_p = \left(e^{\sigma_p^2} - 1\right)e^{2\mu_p + \sigma_p^2} + \left(e^{\tilde{\sigma}_p^2} - 1\right)e^{2\tilde{\mu}_p + \tilde{\sigma}_p^2}$$

A binary variable, $X_p(t)$ can be used to indicate whether patient $p$ is in recovery at time $t$. We assume that the start time of patient p's surgery takes the value $Z_p$.

$$X_p(t) = \begin{cases} 1, & \text{patient } p \text{ is in recovery at time } t \\ 0, & \text{otherwise} \end{cases}$$

$$\Pr(X_p(t) = 1) = \Pr(Z_p + S_p \leq t \leq Z_p + T_p)$$

$$= \Pr(t - Z_p \leq T_p) - \Pr(t - Z_p < S_p)$$

$$= \Pr(S_p \leq t - Z_p) - \Pr(T_p < t - Z_p)$$

$$= F_{S_p}(t - Z_p) - F_{T_p}(t - Z_p)$$

$$= \frac{1}{2}\text{erf}\left(\frac{\ln(t - Z_p) - \mu_p}{\sqrt{2}\sigma_p}\right) - \frac{1}{2}\text{erf}\left(\frac{\ln(t - Z_p) - \hat{\mu}_p}{\sqrt{2}\hat{\sigma}_p}\right),$$

$$Z_p \leq t \leq Z_p + \exp\left(\frac{\hat{\sigma}_p \mu_p - \sigma_p \hat{\mu}_p}{\hat{\sigma}_p - \sigma_p}\right)$$

### 3.2.2 Forecasting the Number of Patients in Recovery

Let $N(t)$ be the number of patients in recovery at time $t$. Here, in order to simplify remaining calculations, we assume that patient surgery durations and recovery times are independent. Based on this assumption, $N(t)$ follows a Poisson Binomial distribution. It is possible to determine the expectation and variance of the number of patients in recovery at time $t$.

$$E[N(t)] = \sum_{p \in P} \Pr(X_p(t) = 1)$$

$$= \sum_{p \in P} \left(\frac{1}{2}\text{erf}\left(\frac{\ln(t - Z_p) - \mu_p}{\sqrt{2}\sigma_p}\right) - \frac{1}{2}\text{erf}\left(\frac{\ln(t - Z_p) - \hat{\mu}_p}{\sqrt{2}\hat{\sigma}_p}\right)\right)$$

$$Var[N(t)] = \sum_{p \in P} \Pr(X_p(t) = 1)\left(1 - \Pr(X_p(t) = 1)\right)$$

$$= \sum_{p \in P} \left(\frac{1}{2}\text{erf}\left(\frac{\ln(t - Z_p) - \mu_p}{\sqrt{2}\sigma_p}\right) - \frac{1}{2}\text{erf}\left(\frac{\ln(t - Z_p) - \hat{\mu}_p}{\sqrt{2}\hat{\sigma}_p}\right)\right)$$

$$\times \left(1 - \frac{1}{2}\text{erf}\left(\frac{\ln(t - Z_p) - \mu_p}{\sqrt{2}\sigma_p}\right) + \frac{1}{2}\text{erf}\left(\frac{\ln(t - Z_p) - \hat{\mu}_p}{\sqrt{2}\hat{\sigma}_p}\right)\right)$$

The cumulative distribution function (**CDF**) of $N(t)$ is determined according to Hong (2013).

$$F_{N(t)}(k) = \frac{1}{n+1}\sum_{l=0}^{n}\frac{\{1 - \exp[-i\omega l(k+1)]\}x_l}{1 - \exp(-i\omega l)},$$

$$\text{where } \omega = \frac{2\pi}{n+1},$$

$$\text{and } x_l = \prod_{j=1}^{n}[1 - p_j + p_j \exp(i\,\omega l)].$$

When we require the evaluation of the above CDF at every value of $t$, it is too computationally expensive for implementation in real-world situations. Here, in order to estimate 95% prediction intervals, we approximate $N(t)$ with a normal distribution, $Y(t)$.

$$N(t) \approx Y(t) \sim N(\text{E}[N(t)], \text{Var}[N(t)])$$

Historical data indicates that this approximation is suitable when examining the 95% prediction intervals.

$$\Pr(L(t) \leq Y(t) \leq U(t)) = 0.95,$$

$$\text{where } L(t) = \text{E}[N(t)] - 1.96\sqrt{\text{Var}[N(t)]},$$

$$\text{and } U(t) = \text{E}[N(t)] + 1.96\sqrt{\text{Var}[N(t)]}.$$

### 3.2.3 Model Validation

We validate the model presented above by comparing forecasts to historical PACU occupancy levels through the duration of 2016. In particular, we are interested in the mean, variance, and estimated 95% prediction intervals. When calculating the historical expected occupancy profile, we use the set of patients who received on that day, and the expected start times of their surgeries. Duration parameters vary by ASA and surgery type, however a lognormal distribution is still a good fit. Figure 1 shows the actual recovery occupancy, the expected recovery occupancy, and the estimated 95% prediction interval for a week in February 2016.

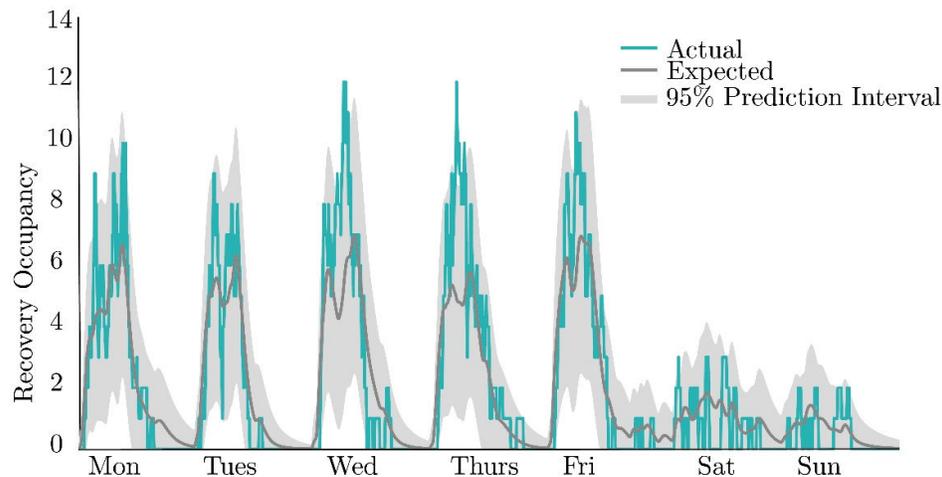

Figure 1: Example occupancy levels of the SCU at the case study hospital based on a week of historical data. The teal line indicates actual occupancy, whilst the grey line represents the expected occupancy calculated using a Poisson Binomial distribution, knowing only the expected start times of surgeries scheduled for each day. The grey shading represents the 95% prediction interval around occupancy levels.

Figure 1 indicates that there is a good agreement between the model and historical data. The peaks and troughs in historical PACU occupancy align well with the peaks and troughs in expected occupancy. The historical recovery occupancy is within the shaded 95% prediction interval for most of the week.

When validating the model against historical data, we calculate the historical and expected occupancies every 0.1 hours for the duration of 2016. On average, there was a difference of 0.07 in the historical and expected occupancy levels. As such, the model presented here slightly underestimates occupancy levels. The model overestimates occupancy 49.5% of the time and underestimates 49.6% of the time. Approximately 0.9% of the time, the model and historical data were in exact agreement. Historical data showed actual occupancy exceeded the upper bound on the approximate 95% prediction interval 3.3% of the time. Actual occupancy was below the lower bound on the approximate 95% prediction interval only 0.2% of the time.

In future it may be worth considering modelling durations phase-type distributions rather than lognormal distributions (e.g. Fackrell (2008)). It may also be possible to refine the probabilistic forecast by incorporating real-time schedule realisations to assist in the recovery occupancy predictions. This is not included in the model as we focus instead on start-of-day schedules, prior to any schedule realisations.

### *3.3 Surgical case sequencing model*

In this section, we present a multi-objective MINLP formulation for the surgical case sequencing problem. The maximum expected number of patients in recovery at any one time is minimised. As such, this model levels the occupancy of the PACU.

#### 3.3.1 Scalar Parameters

$\bar{\bar{H}}$: the number of surgeons scheduled for the day.

$\bar{P}$: the number of patients scheduled for the day.

$\bar{R}$: the number of ORs.

$\lambda$: the number of hours that each OR is open for during a standard working day.

$\lambda^*$: the number of hours in a day.

### 3.3.2 Index Sets

$H$: the set of surgeons that practice at the hospital. $H = \{1, \ldots, \bar{H}\}$

$P$: the set of patients that are scheduled for the day. $P = \{1, \ldots, \bar{P}\}$

$R$: the set of ORs. $R = \{1, \ldots, \bar{R}\}$

$P_h$: the set of patients treated by surgeon $h, \forall h \in H$.

$P_r$: the set of patients treated in OR $r, \forall r \in R$.

### 3.3.3 Indices

$h$: index for surgeon in set $H$.

$p$: index for patient in set $P$.

$q$: alternative index for patient in set $P$.

$r$: index for OR in set $R$.

### 3.3.4 Vector Parameters

$\beta_h$: the setup time surgeon $h$ requires when starting in a new OR, $\forall h \in H$.

$\gamma_p$: 1 if patient p requires post-surgery recovery in the recovery space, 0 otherwise, $\forall p \in P$.

$\tau_p$: the expected duration of patient $p$'s surgery, $\forall p \in P$.

$\mu_p$: the mean of the natural logarithm of patient $p$'s surgical duration, $\forall p \in P$.

$\sigma_p^2$: the variance of the natural logarithm of patient $p$'s surgical duration, $\forall p \in P$.

$\tilde{\mu}_p$: the mean of the natural logarithm of patient $p$'s recovery duration, $\forall p \in P$.

$\tilde{\sigma}_p^2$: the variance of the natural logarithm of patient $p$'s recovery duration, $\forall p \in P$.

$\hat{\mu}_p$: the mean of the natural logarithm of the duration of patient $p$'s surgery and recovery, $\forall p \in P$.

$\hat{\sigma}_p^2$: the variance of the natural logarithm of the duration of patient $p$'s surgery and recovery, $\forall p \in P$.

$\rho_h$: the start of shift time of surgeon $h, \forall h \in H$.

$T_p^+$: the setup time (in hours) before patient $p$'s surgery, $\forall p \in P$.

$T_p^-$: the clean-up time (in hours) after patient $p$'s surgery, $\forall p \in P$.

### 3.3.5  Decision Variables

$V_{pq}$: 1 if patient $q$'s surgery starts during patient $p$'s surgery, 0 otherwise, $\forall p, q \in P$.

$U_{pq}$: 1 if patient $q$'s surgery starts after patient $p$'s sugery starts, 0 otherwise, $\forall p, q \in P$.

$Z_p$: the expected start time of patient $p$'s surgery, $\forall p \in P$.

$Z_p^*$: the expected end time of patient $p$'s surgery, $\forall p \in P$.

$\Omega_h$: 1 if surgeon $h$ will require overtime, 0 otherwise, $\forall h \in H$.

$O_h$: the expected overtime associated with surgeon $h, \forall h \in H$.

### 3.3.6  Objective Function

The main objective is to minimise the maximum expected number of patients in recovery at any one time throughout the surgical department opening hours. In doing so, we level the demand on PACU staff. This not only assists in staff planning but may also result in increased staff satisfaction and better care for patients in recovery. We provide an example occupancy profile in Figure 2.

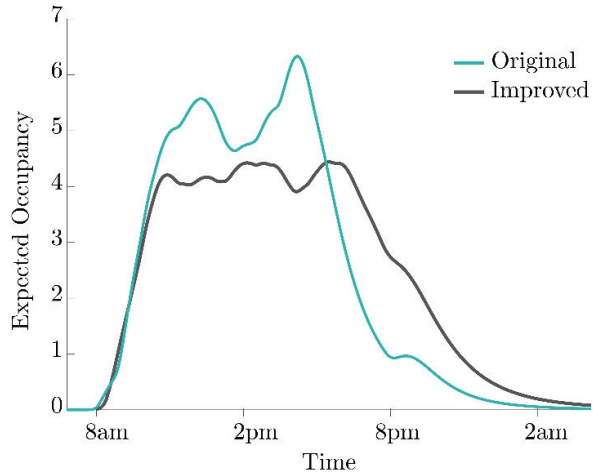

Figure 2: An example occupancy profile of the SCU at the case study hospital. The teal line indicates the expected SCU occupancy over a single day, considering the historical surgery schedule. The grey line indicates the expected occupancy of the SCU on the same day, with the same surgeries, after the schedule has been optimised using the methodology presented in this paper.

Figure 2 includes both the original (historical) expected occupancy profile, and the improved expected occupancy profile. The improved expected occupancy profile is the expected occupancy calculated over the course of a day where the expected start times of surgery are $Z_p$, as determined through this model (solved prior to the opening of the OT).

In this instance, there were 61 patients scheduled for treatment in 21 different ORs by 35 different surgeons. Of the 61 patients scheduled for the day, only 45 required recovery in the PACU. The methodology presented here reduces the maximum expected occupancy (**MEO**) from 6.33 patients to 4.44 patients by making simple adjustments to scheduled surgery start times and patient order.

Whilst the historical occupancy profile exhibits peaks and troughs in demands, by reducing the MEO throughout OT opening hours, the improved occupancy profile is steadier and forecasts level demand throughout opening hours.

We define the objective function based on the work presented in Section 3.2.

Minimise

$$\max_{0 \leq t \leq \lambda^*} \sum_{p \in P} \gamma_p \begin{cases} \frac{1}{2}\text{erf}\left(\frac{\ln(t-Z_p)-\mu_p}{\sqrt{2}\sigma_p}\right) - \frac{1}{2}\text{erf}\left(\frac{\ln(t-Z_p)-\hat{\mu}_p}{\sqrt{2}\hat{\sigma}_p}\right), & Z_p \leq t \leq Z_p + \exp\left(\frac{\hat{\sigma}_p \mu_p - \sigma_p \hat{\mu}_p}{\hat{\sigma}_p - \sigma_p}\right) \\ 0, & \text{otherwise} \end{cases}$$

This objective is equivalent to

Minimise

$$\max_{0 \leq t \leq \lambda^*} \sum_{p \in P} \frac{\gamma_p}{2} \left[\text{erf}\left(\frac{\ln(\max(t-Z_p,0))-\mu_p}{\sqrt{2}\sigma_p}\right) - \text{erf}\left(\frac{\ln(\max(t-Z_p,0))-\hat{\mu}_p}{\sqrt{2}\hat{\sigma}_p}\right)\right], \quad (1)$$

when $Z_p + \exp\left(\frac{\hat{\sigma}_p \mu_p - \sigma_p \hat{\mu}_p}{\hat{\sigma}_p - \sigma_p}\right) > \lambda^*$. This condition holds for the parameters of interest in the case study.

### 3.3.7 Constraints

Surgeon $h$'s surgeries must start after the start of their shift.

$$Z_p \geq \rho_h, \forall\ h \in H, p \in P_h \quad (2)$$

Constraint (3) is used to ensure that surgeries are completed before the end of surgeon $h$'s shift.

$$Z_p + \tau_p \leq \rho_h^* + O_h, \forall\ h \in H, p \in P_h \quad (3)$$

In some instances, overtime may be required (as seen in constraint (3)). For each surgeon, the amount of overtime available is limited by the expected duration of their patient's surgeries, with provision for setup and cleanup times. In most cases, surgeons will not be allocated overtime.

$$O_h \leq \Omega_h \left(\sum_{p \in P_h}(\tau_p + T_p^+ + T_p^-) - \rho_h + \rho_h^*\right), \forall\ h \in H \quad (4)$$

Constraints (5) and (6) determine whether patient $q$ ends on or after the time patient $p$ starts.

$$Z_p - Z_q^* > -\lambda^* U_{pq}, \forall p, q \in P, p \neq q \qquad (5)$$

$$Z_q^* - Z_p \geq \lambda^*(U_{pq} - 1), \forall p, q \in P, p \neq q \qquad (6)$$

Constraints (7) and (8) determine whether the treatment of patients overlaps. Although we could use a single constraint, that constraint would be nonlinear.

$$V_{pq} \geq U_{pq} + U_{qp} - 1, \forall p, q \in P, p \neq q \qquad (7)$$

$$2V_{pq} \leq U_{pq} + U_{qp}, \forall p, q \in P, p \neq q \qquad (8)$$

Each surgeon must not treat more than one patient at a time.

$$V_{pq} = 0, \forall h \in H, p, q \in P_h, p \neq q \qquad (9)$$

If two patients' surgeries overlap, then they must not use the same OR.

$$V_{pq} = 0, \forall r \in R, p, q \in P_r, p \neq q \qquad (10)$$

We use constraint (11) to calculate the expected finish time of each surgery.

$$Z_p^* = Z_p + \tau_p, \forall p \in P \qquad (11)$$

If we treat two patients with the same surgeon, or in the same OR, then they require clean-up and setup time between surgeries.

$$Z_q \geq Z_p^* + T_q^+ + T_p^- - \lambda(1 - U_{pq}), \forall h \in H, p, q \in P_h, p \neq q \qquad (12)$$

$$Z_q \geq Z_p^* + T_q^+ + T_p^- - \lambda(1 - U_{pq}), \forall r \in R, p, q \in P_r, p \neq q \qquad (13)$$

Overtime must be non-negative.

$$O_h \geq 0, \forall h \in H \qquad (14)$$

### 3.3.8 Model Assumptions

We make a number of assumptions to simplify the model presented:

- Surgical and recovery durations are independent.

- We do not consider non-elective patients as the hospital currently has other reservation techniques for their surgeries.

- Each patient's surgeon and OR is known in advance based on the surgical case assignments made on a weekly basis.

- There is no delay between a patient's surgery and recovery.

- Overtime may be unavoidable depending on the predetermined surgical case assignments.

## *3.4 Solution Approach*

The SCSP is NP-hard as it is reducible to a machine scheduling problem. In addition to this, the objective function presented in Section 3.3.6 is nonlinear, nonconvex, and not continuously differentiable. Commercial solvers are inadequate.

Simulated Annealing (**SA**) is a simple metaheuristic based on the cooling of metals (Kirkpatrick, Gelatt, & Vecchi, 1983). By occasionally accepting worsening solutions, SA is able to escape local optima in the search of global optima. Under certain conditions, SA converges to a solution arbitrarily close to the global optimum with a probability arbitrarily close to 1 (Mitra, Romeo, & Sangiovanni-Vincentelli, 1986).

Baesler, Gatica, and Correa (2015) compare SA to a number of common sequencing heuristics on an OR scheduling problem with the objective of minimising makespan. The authors found that SA significantly outperformed each of the heuristics tested (shortest processing time, longest processing time and first-in first-out). Additionally, the simulation optimisation approach implemented produced schedules 18% better than the hospital's current schedules.

SA is a reasonably intuitive metaheuristic which can be understood at a broad level by hospital staff unfamiliar with metaheuristic approaches. Given the simplicity of the algorithm, ease of implementation, solution quality, and computational efficiency, SA solves the SCSP presented in Section 3.3. We implement the metaheuristic in MATLAB R2017b on a desktop computer with an Intel® Core™ i7-6700 CPU with 16GB of RAM.

When implementing SA, it is necessary to define local search heuristics to explore solution neighbourhoods. Here, we exploit the problem structure in order to find local neighbours. At each iteration, we swap the order of two random patients, then Algorithm 1 (cf. Section 3.4.1) randomly generates a schedule based on the patient order. SA always accepts a sequence that reduces the MEO of the PACU (i.e. reduces the objective function value) and occasionally accepts a solution that increases the MEO of the PACU. In doing so, the metaheuristic is able to escape local optima in the search for global optima. The 'temperature' of SA is related to how likely it is that the algorithm will accept a worsening solution. This temperature reduces over time such that we prioritise breadth of search initially, and depth of search is the priority towards the end of the computational time.

### 3.4.1 Constructive Heuristic

While using SA, it is necessary to have a local search strategy to determine the expected start times of patients' surgeries given a sequence of patients recommended by SA. As such, at each iteration, we construct a schedule by defining the order of patient's surgeries, and determining the earliest start and latest completion of each patient's surgery based on this order. We allocate these surgeries such that they start after their earliest start, and finish before their latest completion times, unless overtime is necessary. This constructive heuristic is based on the critical path method (Kelley Jr, 1961), with the inclusion of random starting times.

Algorithm 1 contains the pseudocode.

In general, the latest completion time for a patient is the closing time of the OR, whilst the earliest start time is the opening time of the OR. The algorithm considers patients in the reverse order of the sequence proposed by SA. Initially, we compile a list of successors for the patient considered. These successors are those patients who are later in the sequence, and who are treated by the same surgeon, or in the same OR.

If a patient does have successors, we modify their latest completion time by considering the time that it would take to treat the successor patients (incorporating setup and clean-up times). When doing so, we also consider the latest completion of the successor patients.

After modifying the latest completion times for each patient, we now consider the patient sequence in the order recommended by SA. For each patient we compile a list of predecessors. If a patient has predecessors, we modify their earliest start time by considering the time it takes to perform the surgery of the predecessor patient (including setup and clean-up).

Now that all patients have an earliest start and latest completion time, we randomly allocate a surgical start time to each patient. When allocating random times, we ensure that a patient's surgery does not start before the earliest possible start time, and their completion does not occur after the latest possible completion time.

Algorithm 1: Constructive Heuristic.

---

Latest Completion (LC) ← OR closing time

Earliest Start (ES) ← OR opening time

**FOR** p in reverse patient sequence

Successors (S) ← the set of patients after patient p in the same OR or treated by the same surgeon.

**IF** Successors exist

LC(p) ← min(LC(S)-Duration(S)-Setup(S))-Cleanup(p)

**END**

**END**

**FOR** p in patient sequence

Predecessors (P) ← the set of patients before patient p in the same OR or treated by the same surgeon.

**IF** Predecessors exist

ES(p) ← max(ES(P)+Duration(P)+Cleanup(P))+Setup(p)

**END**

Start(p) ← **ES(p) + max(0, rand** ×(LC(p)-ES(p)-Duration(p)))

ES(p) ← Start(p)

**END**

---

### 3.4.2 Parameter Tuning

To obtain good solutions through SA, metaheuristic parameter tuning is required. We use a parameter sweep to determine suitable parameters. We varied the total number of iterations between 1000 and 3000. We considered proportional temperature decreases of 0.85, 0.90 and 0.95 every 50, 100 and 200 iterations. These temperature decreases refer to the way SA accepts a solution with reducing probability over time. We performed parameter tuning in parallel using MATLAB® R2017b on the university's High Performance Computing (**HPC**) facility.

To simplify the parameter tuning, the starting temperature was set to one. SA ran ten times under each parameter combination on historical data from February 2016. Ten runs was sufficient to observe metaheuristic performance. We used the average of the sum of MEO on each day to compare against other parameter combinations. Based on these computational experiments, the metaheuristic performs best with 2500 iterations and a temperature reduction of 5% every 200 iterations.

## 4    Results and Discussion

In this section, we perform computational experiments to compare the actual (historical) MEO to the improved MEO for each day in 2016. We selected metaheuristic parameters based on the parameter tuning performed in Section 3.4.2. Computational experiments used MATLAB® R2017b on a desktop computer with an Intel® Core™ i7 processor @ 3.40GHz with 16GB of RAM.

For each working day in 2016 (i.e. any day that elective surgeries were performed), SA was run ten times. Ten runs was sufficient to observe the behaviour of the metaheuristic under the given parameters. Figure 3 shows the historical MEO compared to improved occupancy for each day in 2016. The teal line shows a cubic spline through the data.

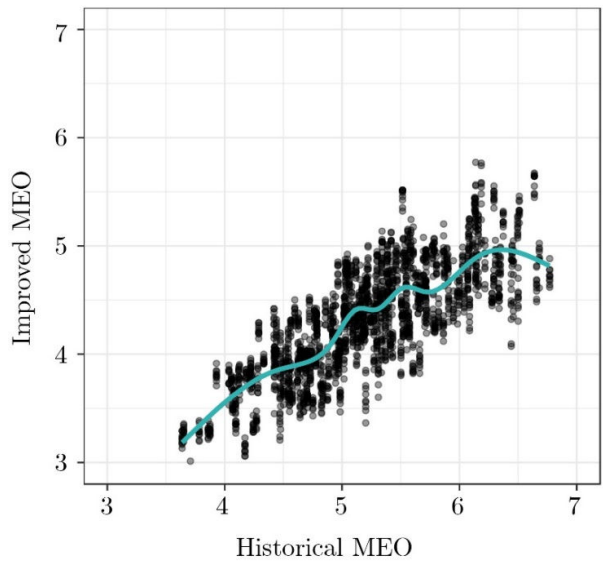

Figure 3: Changes to the MEO in the SCU at the case study hospital. On the x-axis we plot the MEO of the SCU as calculated using the historical surgical schedule. On the y-axis we plot the MEO of the SCU after the methods in this paper have been used to modify the surgical sequences. A spline is fit to show trends. This plot shows all data from 2016.

Figure 3 shows that whilst there are a few instances where this approach did not improve historical MEO, in most instances, our approach decreased the MEO significantly. On average, there was an 18% reduction in MEO. In some instances this reduction was as large as 47%. Such large reductions in MEO have the effect of levelling the workload of staff in the PACU. This, in turn, may result in improved patient care and reduced levels of stress in staff members.

Figure 4 shows how the throughput of the OT (in number of elective patients treated) affects the MEO of the PACU. We show historical throughput on each day in 2016 against to the historical (black) and improved (teal) MEO of the PACU on that day. A cubic spline shows the general trend. We see larger relative reductions in MEO as the number of elective patients treated increases. In both the historical and improved case, the MEO is approximately proportional to the number of elective patients treated. The exception to this is the slight downward trend in improved MEO seen at around 50 patients. It is unclear whether this is a true trend, or if this is an artefact of having few observations at those values of throughput.

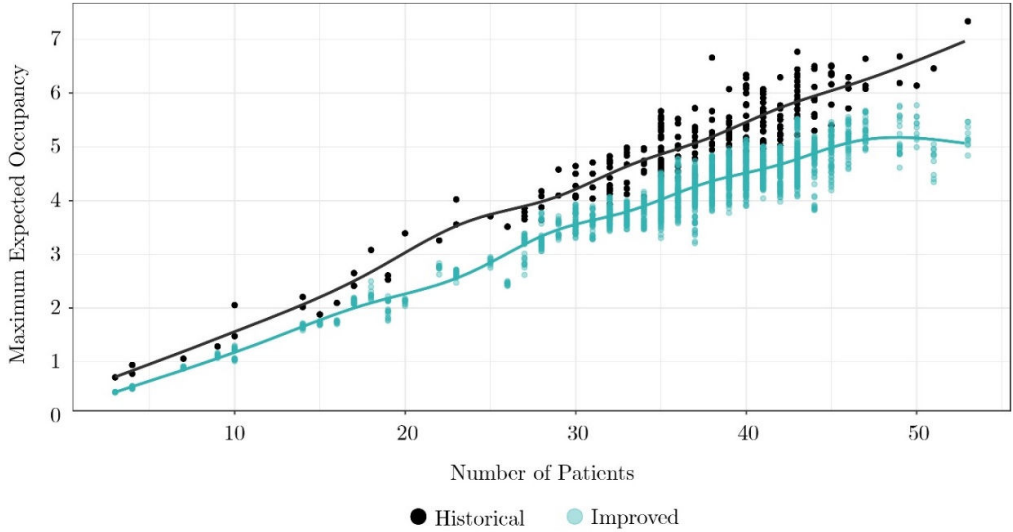

Figure 4: Throughput compared to MEO. MEO of the SCU at the case study hospital is plotted against the patient throughput of the OT. The historical performance is shown in black, whilst the improved performance (in teal) demonstrates that the MEO of the SCU after using the methods in this paper. In each case, a spline is fit to show trends. Data from each day in 2016 is used.

The metaheuristic approach presented in Section 3.4 runs in time $O(n)$, where $n$ is the number of elective patients to be sequenced on a given day. We performed computational experiments using MATLAB R2017b on a desktop computer with an Intel® Core™ i7-6700 CPU with 16GB of RAM. The metaheuristic approach requires under six seconds to run on a standard desktop computer. This ensures that the solution methodology is accessible to hospital planning staff. Given such a short runtime, it would be reasonable to produce several schedules for the day and allow staff members to use their expert knowledge to select the most appropriate schedule.

## 5 Concluding Remarks

In this paper we presented a probabilistic start-of-day forecast for the occupancy of the post-surgery recovery spaces. This model uses lognormally distributed surgery and recovery durations. The only inputs required to forecast recovery bed occupancy are the expected start times of each patient's surgery, and the parameters associated with the patients' surgery and recovery duration distributions. Validation on historical data shows that this model is effective at capturing the mean recovery bed occupancy levels.

Based on the recovery occupancy model, we provide a MINLP formulation for the SCSP. The objective is to minimise the maximum expected recovery bed occupancy. Given the complexity of the SCSP, and the intricacies of the objective function, it is necessary to implement metaheuristic approaches to produce good feasible solutions. We reduce the maximum expected recovery bed occupancy by 18% on average, compared to historical data at the case study hospital.

If surgeons and administrators have particular constraints around patient order or preferences, minor modifications will incorporate these in the model. As such, we propose that a decision support tool could incorporate this method to assist in the OT planning and scheduling process. Additionally, hospital staff could use probabilistic forecast of post-surgery recovery occupancy levels as a stand-alone tool.

The work presented in this paper will ensure more level demand on the PACU. This may have the effect of reducing stress on PACU staff, improving the quality of care for patients, and reducing bottlenecks associated with this limited resource.

## 6 Acknowledgements

This research was funded by the Australian Research Council (**ARC**) Linkage Grant LP 140100394. Computational resources and services used in this work were provided by the High

Performance Computing and Research Support Group, Queensland University of Technology, Brisbane, Australia.

# 7 Declaration of Interests

The authors declare they have no competing interests.